\documentclass[letterpaper,floatfix,aps,pra,amsmath,amssymb,twocolumn,superscriptaddress,nopacs]{revtex4-1}

\usepackage{graphicx}
\usepackage{verbatim}
\usepackage{color}
\usepackage{array}
\usepackage[pdfauthor={Riwar, Catelani},
            pdftitle={Efficient quasiparticle traps with low dissipation through gap engineering}]{hyperref}

\hypersetup{pdfstartview={XYZ null null 1.05}}

\newcommand{\be}{\begin{equation}}
\newcommand{\ee}{\end{equation}}
\newcommand{\bea}{\begin{eqnarray}}
\newcommand{\eea}{\end{eqnarray}}

\begin{document}

\title{Efficient quasiparticle traps with low dissipation through gap engineering}

\author{R.-P. Riwar}

\affiliation{JARA Institute for Quantum Information (PGI-11), Forschungszentrum J\"ulich, 52425 J\"ulich, Germany}

\author{G. Catelani}

\affiliation{JARA Institute for Quantum Information (PGI-11), Forschungszentrum J\"ulich, 52425 J\"ulich, Germany}

\begin{abstract}
Quasiparticles represent an intrinsic source of perturbation for superconducting qubits, leading to both dissipation of the qubit energy and dephasing. Recently, it has been shown that normal-metal traps may efficiently reduce the quasiparticle population and improve the qubit lifetime, provided the trap surpasses a certain characteristic size. Moreover, while the trap itself introduces new relaxation mechanisms, they are not expected to harm state-of-the-art transmon qubits under the condition that the traps are not placed too close to extremal positions where electric fields are high. Here, we study a different type of trap, realized through gap engineering. We find that gap-engineered traps relax the remaining constraints imposed on normal metal traps. Firstly, the characteristic trap size, above which the trap is efficient, is reduced with respect to normal metal traps, such that here, strong traps are possible in smaller devices. Secondly, the losses caused by the trap are now greatly reduced, providing more flexibility in trap placement. The latter point is of particular importance, since for efficient protection from quasiparticles, the traps ideally should be placed close to the active parts of the qubit device, where electric fields are typically high.
\end{abstract}

\date{\today}


\maketitle

\section{Introduction}
Superconducting circuits are prime candidates for a successful physical implementation of qubits, the fundamental building blocks of quantum computers~\cite{sqr,koch2007,Manucharyan_2009}. In order to ensure a stable and sufficiently error-free computation, qubits need to be protected from possible perturbations. In particular, quasiparticle excitations pose a serious problem as an intrinsic source of errors~\cite{lutchyn,prl,leppa,shaw,martinis,3dtr,riste}. Their harmful effect comes from the coupling to the qubit degrees of freedom, when they tunnel across a Josephson junction, as was shown in~\cite{prb1}. This mechanism can be a particularly severe issue, because experiments have shown that quasiparticles occur at much higher densities than those expected in thermal equilibrium~\cite{martinis,riste}.

In order to mitigate the negative effects of quasiparticles, one needs to find ways to evacuate them from the active parts of the device. So far two classes of strategies have been proposed and tested. The first class, the one we focus on in this paper, includes introducing local regions containing subgap states into which quasiparticles can relax. Such quasiparticle traps have been realized through introducing vortices~\cite{ullom,plourde,wang,pekola2}, tunnel coupling the device to normal metals~\cite{court,raja1,raja2,Riwar,Patel_2016,trapopt}, or through gap engineering~\cite{cpt,sun}. A second strategy involves a time-dependent control of the device, in order to pump quasiparticles away through pulses~\cite{Gustavsson1573}.

In the present paper, we theoretically study traps implemented via gap engineering -- that is, by coupling the superconductor $S$ used to fabricate the active device part (usually, one or more Josephson junctions) to a different superconductor $S'$ with a lower gap. The effect of gap engineering was extensively studied for Cooper-pair transistors~\cite{cpt,Ferguson2006,Naaman2006,Shaw2008,Court2008}. In essence, the circuit can be efficiently protected from quasiparticles by making transistors where the central islands have a higher gap than the leads. For circuits with only single junctions, such as transmons, the gap engineering must performed in a more sophisticated manner, e.g., by proximitizing parts of the circuit. This was attempted in~\cite{sun}, where the gap was locally increased at the junction; however there was no unequivocal proof of a beneficial effect, possibly due to a non-optimal choice of the design. Here, we propose different designs of gap-engineered traps located further away from the junction, building on our previous insights from normal metal traps, which were shown to be efficient~\cite{Riwar}. In particular, we predict an improvement compared to normal metal traps, both in terms of  increased trapping efficiency and reduction of unwanted adverse effects.

In order to appreciate these improvements, we summarize our past works on normal metal traps~\cite{Riwar,trapopt,Riwar2018}. The most important conclusions are the following. There is a minimal trap size above which the trap is ``strong''~\cite{Riwar}, in the sense that the quasiparticle evacuation rate is limited by diffusion through the device. Importantly, this minimal trap size depends on the quality of the superconductor-normal metal ($SN$) interface, and can become large due to a quasiparticle backflow stemming from the peak in the superconducting  density of states just above the gap. This limits the effectiveness of normal-metal traps in small devices. For sufficiently large traps, the quasiparticle evacuation can be significantly accelerated by splitting the trap into many pieces, and distributing them evenly over the device~\cite{trapopt}. However, likewise the positive effect of trap splitting is limited, as soon as the individual trap pieces approach the minimal size. Finally, we have shown that placing traps close to active parts of the circuit is favorable for the qubit lifetime, as it suppresses the quasiparticle density and its fluctuations at the place where it does the most harm.

On the other hand, normal metal traps themselves introduce new processes that could limit the quality factor of the qubit~\cite{Riwar2018,Hosseinkhani2018Feb}. The two most important processes are the dissipation in the normal metal bulk due to the ac charge redistribution in the circuit, and the losses due to photo-assisted tunneling at the $SN$ interface. While the latter is expected to be sufficiently small, the former process can potentially pose problems if traps are placed too close to active parts of the circuit where electric fields are high. This issue is thus in competition with the optimization principle mentioned above, where traps close to the active parts increases the protection from the nonequilibrium quasiparticles. Normal metals can also induce, via the inverse proximity effect, subgap states that increase the qubit decay rate, but this effect is exponentially suppressed with distance from the active parts and can be neglected.

We here propose to replace the weakly-coupled normal metal pieces, with strongly coupled superconducting parts with a significantly lower gap. For sufficiently good interfaces, the proximity effect will introduce a spatially dependent gap in the circuit (see e.g.~\cite{Belzig1999May,Hosseinkhani2018Feb} and references therein) and provide a region where the quasiparticles can be trapped. Studying the quasiparticle diffusion dynamics, we predict that the minimal trap size to reach the strong trapping limit is reduced with respect to normal metal traps, making efficient traps possible in devices of smaller sizes. In particular, the unfavorable backflow of quasiparticles from normal metals, is absent here. Secondly, we investigate the dominant dissipation processes induced by the gap-engineered traps. We find that charge redistribution currents no longer pose a threat, since the presence of the lower gap reduces the bulk resistivity by several orders of magnitude. Photo-assisted tunneling processes on the other hand can now become more important since the improved interfaces increase the tunneling rate. Overall, we show however that in practice gap-engineered traps do not cause any harmful side effect.

This paper is structured as follows. In Sec.~\ref{sec_diff_and_trapping} we present the diffusion model for qubits with gap-en\-gineered traps and provide our results for the quasiparticle diffusion dynamics. In Sec.~\ref{sec_dissipation} we estimate the dominant dissipation processes induced by these traps. Our conclusions are presented in Sec.~\ref{sec_conclusions}. In Appendix~\ref{app:diffusion} we provide a microscopic justification for our diffusion model. In Appendix~\ref{app:quantum_circuit_dissipation} we detail our estimate for the dissipation due photo-assisted tunneling. Finally, in Appendix~\ref{app:coplanar} we compute the geometric capacitances for coplanar circuits.

\section{Diffusion and trapping of quasiparticles in an $SS'$ trap}
\label{sec_diff_and_trapping}

\subsection{The diffusion model}

\begin{figure}[t]
\begin{center}
\includegraphics[width=0.65\columnwidth]{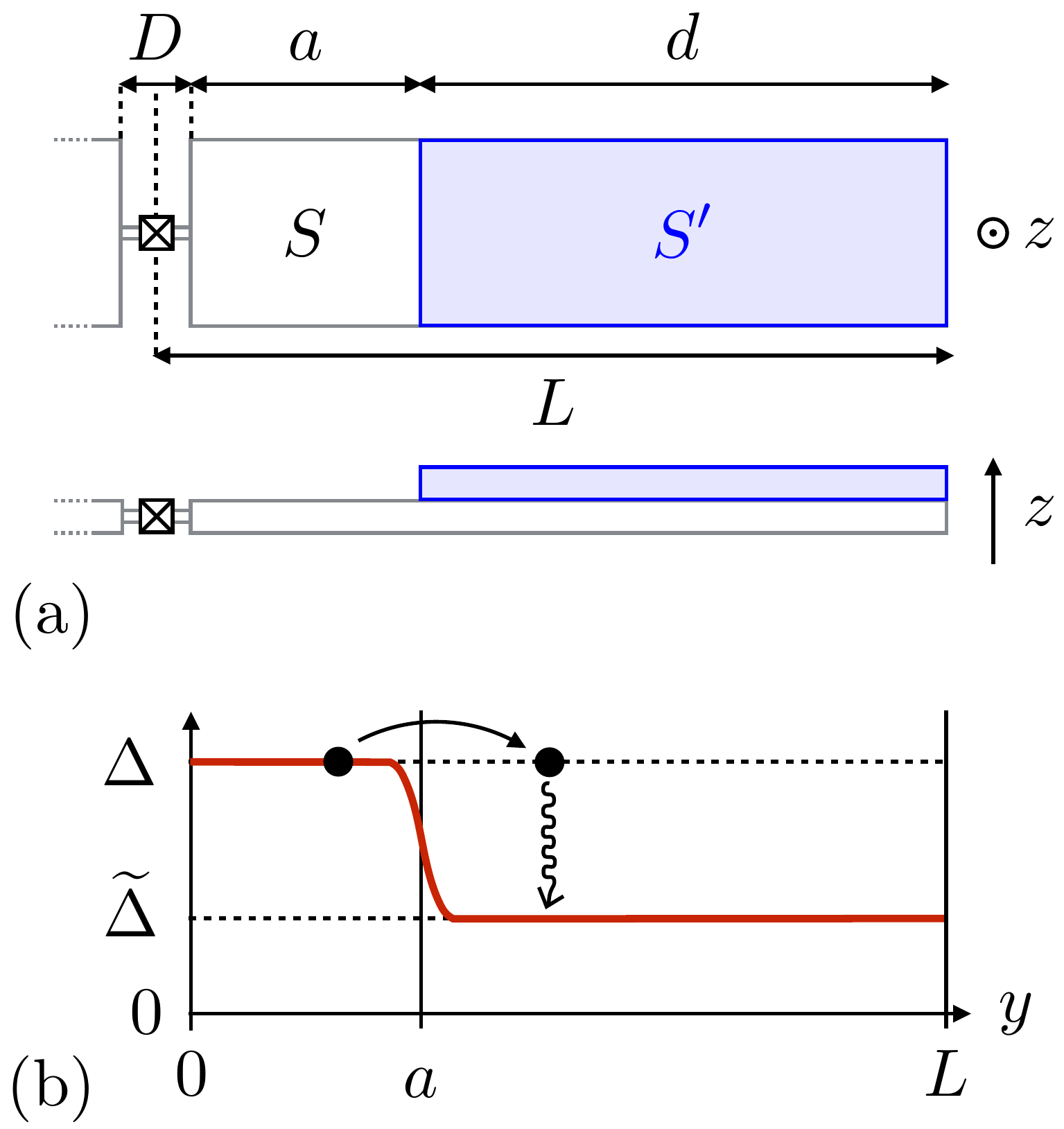}
\end{center}
\caption{(a) Transmon qubit made out of superconductor $S$ with a gap $\Delta $. The trap is realized through gap engineering with a second superconductor $S'$ which has a lower gap $\Delta_{S'}$. The middle dashed line marks the origin of the $y$ axis (horizontal direction). One transmon side has the total length $L$. The second superconductor $S'$ is attached to the transmon with a distance $a$ from edge to edge, and has length $L'$. The two transmon plates are separated by a small distance $D\ll a,\,L$. For most of the discussion (unless specified otherwise) we set $D\rightarrow 0$. (b) The resulting position dependence of the gap inside the transmon, as a function of $y$. The change of the gap parameter is occurring on the length scale of the coherence length. Once the quasiparticles diffuse into the region with the lower gap, they may relax to energies below $\Delta$, and are thus trapped.}
\label{fig_SSp_model}
\end{figure}

We study the following model. The starting point consists of a conventional transmon circuit, made from a certain superconducting material $S$ with gap $\Delta$. Part of this circuit is in contact with a second superconductor $S'$, with a lower gap $\Delta_{S'}$, see Fig.~\ref{fig_SSp_model}a. We are interested in the limit, where there is a good contact between the two superconductors, such that $S$ gets proximitized, making its gap position-dependent, $\Delta(y)$  (see Fig.~\ref{fig_SSp_model}b). The proximity effect thus provides a gap-engineered quasiparticle trap in the transmon~\footnote{In fact, the same principle could be implemented with a good contact to a sufficiently thin normal metal -- see, e.g., Refs.~\cite{Hosseinkhani2018Feb,Belzig1999May} and references therein.}. Quasiparticles at energy $\approx\Delta$ or above will then, once they diffuse into the lower gap region, relax, and consequently become confined to the proximitized region. The essential trapping principle is therefore the same as in the previously studied normal-metal traps, except that with proximitized traps we expect several improvements in terms of trapping performance, as well as a reduction of unwanted side effects, as we detail in the course of this paper.

As indicated in Fig.~\ref{fig_SSp_model}b, deep within the the proximitized region, the gap approaches a certain value $\widetilde{\Delta}$, with $\Delta_{S'}\le \widetilde{\Delta} \le \Delta$.
The smooth change from $\Delta$ to $\widetilde{\Delta}$ in the proximitized part of the superconductor occurs on the scale of the coherence length~\footnote{Since the coherence length is much larger than the Fermi wavelength (in low critical-temperature superconductors), we can neglect quasiparticle scattering events across this interface.}.

We will show below that the coherence length ($\sim 100\,$nm in thin aluminum films~\cite{Steinberg2008,Catelani2008}) is small in comparison to the diffusion length scales, such that on the level of the diffusion equation, we can assume an immediate, discontinuous drop of the gap parameter $\Delta\left(y\right)=\Delta\theta\left(a-y\right)+\widetilde{\Delta}\theta\left(y-a\right)$, for $0\leq y\leq L$, where $L$ is the total device size, and $a\leq y\leq L$ is the proximitized part.

Finally, in the following, we are interested in the quasiparticles located in the active parts of the circuit, which can actually disrupt the qubit state.  This concerns the quasiparticles at energies above $\Delta$, for which we define the dimensionless density (normalized by the density of Cooper pairs)
\begin{equation}\label{eq_xqp_def}
x_{\text{qp}}\left(y\right)=\frac{2}{\Delta}\int_{\Delta}^{\infty}d\epsilon\frac{\epsilon}{\sqrt{\epsilon^{2}-\Delta^{2}\left(y\right)}}f_{\text{qp}}\left(\epsilon,y\right)\ .
\end{equation}
For this quantity, we can formulate the phenomenological diffusion equation
\begin{equation}\label{eq_diffusion_discontinuous}
\dot{x}_{\text{qp}}\left(y\right)=\partial_{y}\left[D_{\text{qp}}\left(y\right)\partial_{y}x_{\text{qp}}\left(y\right)\right]-\tau_{\text{r}}^{-1}\theta\left(y-a\right)x_{\text{qp}}\left(y\right) \ .
\end{equation}
For the microscopic justification of this equation from a more general energy-dependent diffusion term, see Appendix~\ref{app:diffusion}.

In the diffusion equation we have included a relaxation term -- the last term on the right-hand side -- responsible for trapping quasiparticles through relaxation to energies below $\Delta$. In order to compare the proximitized traps to the normal-metal ones, we assume that such a relaxation rate is due {in both cases} to inelastic electron-phonon interactions, and can therefore be computed according to Ref.~\cite{Kaplan1976}. For $\Delta\gg\widetilde{\Delta}$ and temperatures small compared to $\widetilde{\Delta}/k_B$, it results in
\begin{equation}\label{eq_tau_th}
\tau_{\text{r}}^{-1}\approx\left(1-6\frac{\widetilde{\Delta}^2}{\Delta^2}\left[\ln\left(2\frac{\Delta}{\widetilde{\Delta}}\right)-1\right]\right)\tau_{N}^{-1}\ ,
\end{equation}
and approaches the normal-metal relaxation rate $1/\tau_N$~\footnote{By assumption, the rate $\tau_N^{-1}$ depends on the electron-phonon coupling strength; it is related to the rate $\tau_0^{-1}$ of Ref.~\cite{Kaplan1976} by $\tau_N^{-1}=\tau_0^{-1}(\Delta/k_B T_c)^3/3 \approx 1.83\tau_0^{-1}$}.
For almost equivalent gaps, $\Delta\gtrsim\widetilde{\Delta}$, we receive
\begin{equation}
\tau_{\text{r}}^{-1}\approx\frac{64\sqrt{2}}{35}\left(1-\frac{\widetilde{\Delta}}{\Delta}\right)^{\frac{7}{2}}\tau_{N}^{-1}\ ;
\end{equation}
that is, a relaxation rate much smaller than the normal-metal relaxation rate. Note that here we have assumed again sufficiently small temperature, $k_{\text{B}}T \ll \Delta-\widetilde{\Delta}$, even though $\Delta-\widetilde{\Delta}\ll\Delta$. Obviously, it is therefore advantageous to choose a sufficiently low gap and strong proximity effect, to achieve fast relaxation. As a realistic example, we note that good contact can be achieved between aluminum and titanium films, and that the Al to Ti (bulk) critical temperature ratio is about 3. By varying the thicknesses of the two materials, and possibly adding a normal-metal layer (e.g., gold), the critical temperature and hence the gap can be tuned over a wide range~\cite{Echternach1999,Zhao2018}.

\subsection{The dynamics of quasiparticle trapping}

In order to examine the quasiparticle diffusion dynamics, we need to find the eigenvalues of Eq.~\eqref{eq_diffusion_discontinuous}. They can be found when rewriting above diffusion equation separately for each side,
\begin{align}\label{eq_diffusion_left}
\dot{x}_{\text{qp}}\left(y\right)=&D_{\text{qp}}\partial_{y}^{2}x_{\text{qp}}\left(y\right)\quad y<a\ , \\ \label{eq_diffusion_right}
\dot{x}_{\text{qp}}\left(y\right)=&\widetilde{D}_{\text{qp}}\partial_{y}^{2}x_{\text{qp}}\left(y\right)-\tau_{\text{r}}^{-1}x_{\text{qp}}\left(y\right)\quad y>a\ ,
\end{align}
and at the interface $y=a$, the quasiparticle density has to be continuous and satisfy the boundary condition 
\begin{equation}\label{eq_bc_xqp}
\widetilde{D}_{\text{qp}}\left.\partial_{y}x_{\text{qp}}\right|_{y=a+0^{+}}=D_{\text{qp}}\left.\partial_{y}x_{\text{qp}}\right|_{y=a-0^{+}}
\end{equation}
which expresses the conservation of current.
In order to study the quasiparticle diffusion dynamics, we proceed similarly as in Ref.~\cite{Riwar,trapopt}. Namely, both Eqs.~\eqref{eq_diffusion_left} and~\eqref{eq_diffusion_right} have eigenmodes of the form $x_\text{qp}(y<a)\sim e^{-\lambda_k t}\cos(ky)$ and $x_\text{qp}(y>a)\sim e^{-\tilde{\lambda}_{\tilde{k}} t}\cos([L-y]\tilde{k})$, respectively, which decay with the constant rates $\lambda_k = D_\text{qp}k^2$ and $\tilde{\lambda}_{\tilde{k}} = \widetilde{D}_\text{qp}\tilde{k}^2+\tau_\text{r}^{-1}$. The respective cosine forms are chosen such as to take into account the hard wall boundary conditions at $y=0$ and $y=L$. Requiring that these rates are the same for an eigenmode of the \textit{total} system (i.e., for $y$ from $0$ to $L$), we find,
\begin{equation}
\tilde{k}=\sqrt{\frac{D_{\text{qp}}}{\widetilde{D}_{\text{qp}}}k^{2}-\frac{1}{L_{\text{r}}^{2}}}\ ,
\end{equation}
where $L_{\text{r}}=\sqrt{\widetilde{D}_{\text{qp}}\tau_{\text{r}}}$. Taking now into account the boundary condition at $y=a$, Eq.~\eqref{eq_bc_xqp}, we find that this imposes an additional condition on $k$,
\begin{equation}
\frac{1}{L_{\text{r}}k}\tan\left(ka\right)=-\frac{L_{\text{r}}\widetilde{k}}{1+\left(L_{\text{r}}\tilde{k}\right)^{2}}\tan\left(\left[L-a\right]\tilde{k}\right)\ . \label{eq:eigenmodes}
\end{equation}
As a result, this fixes the allowed values of $k$ to discrete values, and consequently, also the eigenvalues of the diffusion equation are discrete, due to the finite system size.
At sufficiently long times, the eigenmode with the lowest $k$ will be the dominant one. In fact, one can show that for the lowest $k$, $\tilde{k}$ is always imaginary, $\tilde{k}=i\kappa$, with $\kappa=\sqrt{1/L_\text{r}^2-D_\text{qp}k^2/\widetilde{D}_\text{qp}}$, such that we can rewrite $-\tilde{k}\tan([L-a]\tilde{k})=\kappa \tanh([L-a]\kappa)$. 

We proceed next by assuming that $\widetilde{D}_{\text{qp}}> D_{\text{qp}}$, meaning that the quasiparticles at energies $E\geq\Delta$ diffuse faster in the part with the lower gap. This is due to the quasiparticle diffusion constant decreasing close to the respective gap (see also Appendix~\ref{app:diffusion}). We then arrive at closed expressions for the slowest density decay rate $\lambda_0$ in the limiting cases shown in Tab.~\ref{tab:decay_rates}.

\begin{table}
\centering%
\begin{tabular}{|c|c|c|}
\hline 
trap size & relaxation limit & diffusion limit\tabularnewline
\hline 
\hline 
$a\ll L-a$ & $\phantom{\Bigl(\Bigr)}\lambda_{0}\approx\tau_{\text{r}}^{-1}\phantom{\Bigl(\Bigr)}$ & $\phantom{\Bigl(\Bigr)}\lambda_{0}\approx\frac{\pi^{2}}{4}\frac{D_{\text{qp}}}{a^{2}}\phantom{\Bigl(\Bigr)}$\tabularnewline large trap
 & $\phantom{\Bigl(\Bigr)}a\ll\sqrt{D_{\text{qp}}\tau_{\text{r}}}\phantom{\Bigl(\Bigr)}$ & $\phantom{\Bigl(\Bigr)}a\gg\sqrt{D_{\text{qp}}\tau_{\text{r}}}\phantom{\Bigl(\Bigr)}$\tabularnewline
\hline 
$a\gg L-a$ & $\phantom{\Bigl(\Bigr)}\lambda_{0}\approx\frac{L-a}{L}\tau_{\text{r}}^{-1}\phantom{\Bigl(\Bigr)}$ & $\phantom{\Bigl(\Bigr)}\lambda_{0}\approx\frac{\pi^{2}}{4}\frac{D_{\text{qp}}}{L^{2}}\phantom{\Bigl(\Bigr)}$\tabularnewline small trap
 & $\phantom{\Bigl(\Bigr)}L-a\ll\frac{D_{\text{qp}}\tau_{\text{r}}}{L}\phantom{\Bigl(\Bigr)}$ & $\phantom{\Bigl(\Bigr)}L-a\gg\frac{D_{\text{qp}}\tau_{\text{r}}}{L}\phantom{\Bigl(\Bigr)}$\tabularnewline
\hline 
\end{tabular}
\caption{The decay rate of the lowest mode for different regimes,
as obtained from Eq.(\ref{eq:eigenmodes}). Here large/small trap refers to the trap length $L-a$ in comparison to the length $a$ of the non-proximitized region, see Fig.~\ref{fig_SSp_model}.\label{tab:decay_rates}}
\end{table}

With these results, we are now equipped to compare gap-engineered traps to normal-metal traps~\cite{Riwar,trapopt}. In the latter case, for typical values of the tunneling rate between superconductor and normal metal, the effective trapping rate of quasiparticles close to the trap
is limited by the relaxation process
in the normal metal, rather than by the tunneling process; the resulting trapping rate is $\sim\sqrt{k_{\text{B}}T/\Delta}\tau_{N}^{-1}$, with $T \ll \Delta/k_B$ the (effective) quasiparticle temperature.
Note that this rate is smaller than $\tau_{N}^{-1}$; this suppression is due to a significant escape of
quasiparticles back into the superconductor, due to the strongly peaked density of states in the superconductor.

Additionally, a minimal value for the normal-metal trap size was identified, above which the trap can evacuate quasiparticles efficiently. In this regime, the evacuation
is limited by the time quasiparticles need to diffuse to the trap.
This minimal size can be written in the form
\begin{equation}
d_{N}\sim\sqrt{\frac{\Delta}{k_{\text{B}}T}}\frac{D_{\text{qp}}\tau_{N}}{L},
\end{equation}
where $L$ is again the size of the superconductor. The above formula
indicates (see also Ref.~\cite{trapopt}) that when making
the devices smaller and smaller, there comes a point where the diffusion-limited regime
can no longer be reached, because it would require $d_{N}>L$. 

Importantly, we can contrast this with the gap-en\-gineered traps considered
here. Due to the perfect interface between the $\Delta$ and $\widetilde{\Delta}$
region, there is no limiting tunneling process. Moreover, there cannot occur any escape processes as for normal metal traps. Therefore, the bottleneck process is necessarily the relaxation of quasiparticles, and its rate is no longer reduced due to escape processes by $\sqrt{k_{\text{B}}T/\Delta}$ as for 
normal-metal traps; indeed, the relaxation rate is given directly by $\tau_{\text{r}}^{-1}$, which is $\approx\tau_{N}^{-1}$
for $\Delta\gg\widetilde{\Delta}$ [cf. Eq.~(\ref{eq_tau_th})]. Consequently, when solving Eq.~(\ref{eq:eigenmodes})
for a small trap, $d\equiv L-a$, with $d\ll L,L_{\text{r}}$, that is, $a\approx L$,
we find that the saturation to the diffusion-limited regime
occurs at a different minimal trap size. We can estimate the latter by equating the results for the density decay rate $\lambda_0$ in the relaxation and diffusion-limited cases, see Tab.~\ref{tab:decay_rates}. We find 
\begin{equation}
d_{S}\sim\frac{D_{\text{qp}}\tau_{\text{r}}}{L}\approx\frac{D_{\text{qp}}\tau_{N}}{L}.
\end{equation}
Crucially, this means that with gap-engineered traps, one can achieve the
diffusion-limited regime in smaller devices than it would be possible
with normal metal traps, since $d_N/d_S\sim\sqrt{\Delta/k_B T} \sim 10$ for typical experimental conditions.

We can generalize the above statement to
an arbitrary ratio between $d$ and $L_{\text{r}}$, where the minimal trap size becomes
\begin{equation}\label{eq_dS}
d_{S}\approx L_{\text{r}}\tanh^{-1}\left(\frac{D_{\text{qp}}}{\widetilde{D}_{\text{qp}}}\frac{L_{\text{r}}}{L}\right).
\end{equation}
The above result can be obtained from Eq.~\eqref{eq:eigenmodes}, by approximating it in the limit $\widetilde{k}\approx 1/L_\text{r}$ and $a\approx L$, such that,
\begin{equation}
Lk \tan(Lk)\approx \frac{\widetilde{D}_\text{qp}L}{D_\text{qp}L_\text{r}}\tanh\left(\frac{d}{L_\text{r}}\right)\ .
\end{equation}
The transition from relaxation to diffusion limit occurs when the right-hand side is $\approx 1$, which is obviously satisfied by $d_S$ in Eq.~\eqref{eq_dS}.

Finally, we note that when making the trap region so large that $L-a>a$,
then the length scale $L-a$ gets replaced by $a$, such that the
transition from weak to strong (diffusion-limited) trapping is no
longer dependent on $L-a$. Instead, the transition occurs at
\begin{equation}
a_{S}=\sqrt{D_{\text{qp}}\tau_{\text{r}}}.
\end{equation}
That is, the diffusion-limited regime is entered when the region with higher gap $\Delta$ becomes longer than the diffusion length scale within that region.
We provide the results for the density decay rate in the various regimes in Table \ref{tab:decay_rates}.

\section{Dissipation in a proximitized trap}
\label{sec_dissipation}

While quasiparticle traps improve the qubit relaxation time, at the same time they may provide new paths for the dissipation of the qubit energy as well as dephasing. For normal metal traps, it has been elaborated~\cite{Riwar2018} that the main mechanisms are either due to ac charge redistribution within the normal metal, or photo-assisted tunneling at the superconductor-trap interface. While in~\cite{Riwar2018} we estimate that both mechanisms are sufficiently low such that they do not limit the
quality factor of the best currently available qubits, they could impose limitations on future improved devices. We now show that we expect both mechanisms to be strongly reduced when using proximitized traps.

\subsection{Dissipation due to ac resistance}

The dissipation due to the ac resistance, when redistributing charges in the circuit, should be strongly reduced: the here considered
gap-engineered traps have a nonzero gap everywhere, bringing the dc dissipation
to zero. There will remain however a residual ac response. We briefly review the ac response of superconductors to estimate the degree
of dissipation reduction. In the ac case~\cite{Mattis_1958,Catelani_2010}, we find the
real part of the conductivity as
\begin{equation}
\frac{\text{Re}[\sigma_{S}\left(\omega\right)]}{\sigma_{N}}=\frac{1}{\omega}\int_{-\infty}^{\infty}d\epsilon\,\nu(\epsilon,\epsilon+\omega)\left[f\left(\epsilon\right)-f\left(\epsilon+\omega\right)\right]\,,
\end{equation}
with $\sigma_N$ the normal-state conductivity and
\begin{equation}
\nu(\epsilon,\epsilon')=\frac{\left|\epsilon\epsilon'+\widetilde{\Delta}^{2}\right|}{\sqrt{\epsilon^{2}-\widetilde\Delta^{2}}\sqrt{\epsilon'^{2}-\widetilde\Delta^{2}}}\ .
\end{equation}
We compute the integral in the limit $T\ll\omega\ll\widetilde\Delta$, in which we find
\begin{equation}\label{eq:resigma}
\frac{\text{Re}[\sigma_{S}\left(\omega\right)]}{\sigma_{N}}\approx\frac{1}{2}\left(\frac{2\widetilde\Delta}{\omega}\right)^{\frac{3}{2}}\tilde{x}_{\text{qp}}\, ,
\end{equation}
where $\tilde{x}_\text{qp}$ is the total normalized quasiparticle density in the proximitized region [obtained by the replacement $\Delta\to\widetilde{\Delta}$ in Eq.~(\ref{eq_xqp_def})]. The left hand side of Eq.~(\ref{eq:resigma}) is small for small quasiparticle densities. Consequently, in order to find the resistivity (that is, the inverse of the conductivity), the imaginary part of the conductivity is 
needed. It accounts for the kinetic inductance and
in the limit $\omega,T\ll\widetilde{\Delta}$, is given by~\cite{Tinkhambook}
\begin{equation}
\frac{\text{Im}[\sigma_{S}]}{\sigma_N}=\frac{\pi\widetilde\Delta}{\omega}\ .
\end{equation}
Based on this, we can find the real part of the resistivity, which
we need to estimate the dissipation due to ac currents,
\begin{equation}
\text{Re}[\rho\left(\omega\right)]=\frac{\text{Re}[\sigma_{S}\left(\omega\right)]}{|\sigma_{S}(\omega)|^{2}}\ .
\end{equation}
Due to $\tilde{x}_{\text{qp}}\ll1$, $\text{Re}[\sigma_{S}]\ll\text{Im}[\sigma_{S}]$,
and therefore we find
\begin{equation}
\text{Re}[\rho\left(\omega\right)]\approx\frac{\sqrt{2}}{\pi^{2}}\sqrt{\frac{\omega}{\widetilde\Delta}}\tilde{x}_{\text{qp}}\rho_{N}\ ,
\end{equation}
which is much smaller than the normal-state resistivity $\rho_{N}=1/\sigma_N$: in the superconducting state, the resistivity is reduced by both the factor $\sqrt{\omega/\widetilde\Delta}<1$ and the factor $\tilde{x}_{\text{qp}}\ll1$. 
In practice we expect $\tilde{x}_{\text{qp}}\lesssim10^{-4}$, since either injection of quasiparticles~\cite{wang} or relatively high temperature ($k_B T/\widetilde{\Delta}>0.1$) are needed to arrive at such densities,
so we can safely deduce that the 
ac-dissipation 
is reduced by at least four orders of magnitude, likely more.

This significant reduction allows for more
flexibility regarding the trap placement. In particular, it enables the positioning of quasiparticle traps closer to the active parts of the circuits, where the evacuation of quasiparticles is crucial. For normal metal traps, this represented a serious limitation. Close to the active circuit parts, the electric fields are high and consequently so are the charge redistribution currents. In fact, for normal metal traps, we found~\cite{Riwar2018} that when placing a trap close to these extreme points, the relaxation due to ac dissipation could give a contribution of the order $10\,\%$ to the quality factor of present-day qubits. With the reduction for gap-engineered traps, we expect that this is no longer a concern, even if qubits relaxation time is further improved.

\subsection{Relaxation due to $S$-$S'$ tunneling}

We now consider the contribution to qubit relaxation from quasiparticle tunneling between superconductors $S$ and $S'$. 
For this purpose, we treat the transmon as a nearly harmonic $LC$ oscillator. Within this picture, the dissipation due to quasiparticle tunneling across the $S$-$S$ Josephson junction can be modelled as a resistive shunt. The resulting qubit decay
rate, in absence of the $S$-$S'$ junction, is proportional
to the $RC$-time of the junction, $\tau_{\text{qubit}}^{-1}\sim\text{Re}\left[Y\left(\omega_{0}\right)\right]/C$.
Here, $Y\left(\omega_{0}\right)$ is the admittance of the $S$-$S$ junction
at the qubit frequency $\omega_{0}$~\cite{prb1} and $C$ is the the corresponding capacitance.

\begin{figure}
\begin{center}\includegraphics[width=0.75\columnwidth]{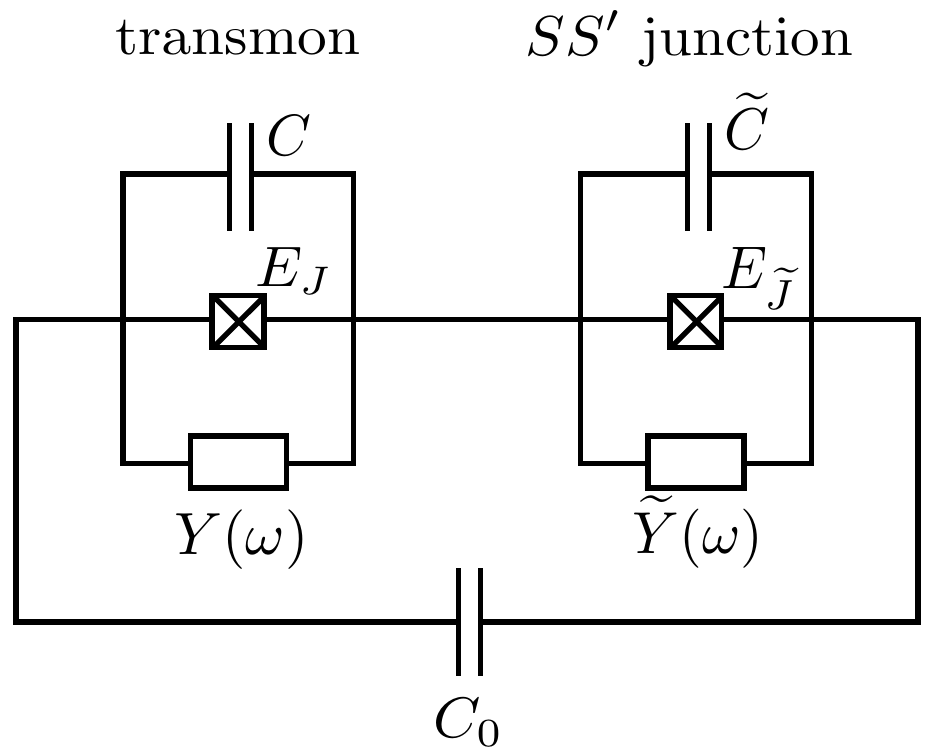}\end{center}

\caption{Dissipative losses induced by the $SS'$ trap. We model both the transmon, as well as the $S$-$S'$ junction in terms of resistively shunted $LC$-circuits. The Josephson energies of the respective junctions are $E_J$ and $\widetilde{E}_J$, and the capacitances are denoted as $C$ and $\widetilde{C}$. The dissipation due to quasiparticle tunneling (whereby energy $\omega$ is dissipated) is described through the real admittances $Y(\omega)$ and $\widetilde{Y}(\omega)$, respectively. The circuits are coupled through the shunting capacitor $C_0$.   \label{fig_dissipative_circuit}}
\end{figure}

Within the same picture, we can now think of the $S$-$S'$ interface as a second $LC$ resonator, coupled to the transmon with a finite capacitive coupling, see Fig.~\ref{fig_dissipative_circuit}. Similarly, quasiparticles can tunnel across the $S$-$S'$ contact, and lead to a resistive shunt; this approach resembles the one used for tunneling at the $N$-$S$ interface of normal-metal traps~\cite{Riwar2018}. As a reminder, in order to effectively trap quasiparticles, we showed before that a good interface is advantageous, to create a trap with a sufficiently low gap. Crucially, the magnitude of the dissipation depends on the interface quality. In fact, there are several competing effects. Obviously, an improved interface increases $Y$ due to the higher tunneling rate, and thus increases the dissipation. Note in particular that the junction conductance can increase by several orders of magnitude, since in the tunneling regime the typical transparency is of order $10^{-5}$, while it is of order one for a good contact. On the other hand, a good tunnel coupling can significantly increase the resonance frequency of the second oscillator, leading to a very small coupling between the two circuits, and thus to a decreasing dissipation.

Actually, a further competition occurs concerning the quasiparticle densities. While a good trap diminishes $x_\text{qp}$ at the junction, and thus increases the $RC$ time of the transmon, the excess quasiparticles will be located within the trap region, thus decreasing the $RC$ time of the $S$-$S'$ junction. We therefore want to ensure that the trap does not become a victim of its own success, by reducing the quasiparticle density at the active device parts to the extent, where the additional dissipation caused \textit{by} the trap could surpass the original dissipative process at the junction. As we will show in the following, this is not a concern for realistic parameters. 

In order to estimate which of the competing processes dominates, we
deploy for the circuit in Fig.~\ref{fig_dissipative_circuit} a simplified classical model.
The equations of motion
for the phase differences across the transmon junction ($\varphi$) and across
the $S$-$S'$ interface ($\tilde{\varphi}$) in frequency-space are
\begin{equation}\label{eq:phivarphi}
\left(M_0-M_\text{diss}\right)\left(\begin{array}{c}
\varphi\left(\omega\right)\\
\tilde{\varphi}\left(\omega\right)
\end{array}\right)=0
\end{equation}
with the matrix $M_0$ describing the 
dissipationless
dynamics of the circuit
\begin{equation}
M_0=\left(\begin{array}{cc}
\frac{C+C_{0}}{2e}\omega^{2}-2eE_{J} & -\frac{C_{0}}{2e}\omega^{2}\\
-\frac{C_{0}}{2e}\omega^{2} & \frac{\widetilde{C}+C_{0}}{2e}\omega^{2}-2e\widetilde{E}_J
\end{array}\right)
\end{equation}
where $E_{J}$ and $C$ are the Josephson energy and capacitance of
the transmon junction, respectively. The tilde designates the same parameters for
the $S$-$S'$ junction. Finally, $C_{0}$ denotes the shunting capacitor coupling
the two junctions. The matrix $M_\text{diss}$ takes into account the dissipative part of the dynamics,
\begin{equation}
M_\text{diss}=\left(\begin{array}{cc}
i\frac{\omega}{e}Y\left(\omega\right) & 0\\
0 & i\frac{\omega}{e}\widetilde{Y}\left(\omega\right)
\end{array}\right)\ .
\end{equation}
Here $Y$ and $\widetilde{Y}$ are the admittances of the corresponding Josephson
junctions, which (as we see later) depend on the junction conductances $g_T$ and $\widetilde{g}_T$, respectively~\footnote{These admittances do not include the parts related to Josephson energy and capacitance, already accounted for by $M_0$, but rather only the real parts of the quasiparticle corrections to those main terms, see Refs.~\cite{prb1} and \cite{Kos2013}.}. The Josephson energies can likewise be related to the respective conductances, $E_{J}=\frac{\Delta}{8}\frac{g_{T}}{g_{0}}$
and $\widetilde{E}_J=\frac{\widetilde{\Delta}}{8}\frac{\widetilde{g}_{T}}{g_{0}}$, where $g_0=e^2/2\pi$ is the conductance quantum~\footnote{We remind that we assume that the phase difference across the $S$-$S'$ junction is sufficiently small that it can be treated within linear response; then the current-phase relation for short junctions is the same for any transparency, see e.g. Ref.~\cite{Beenakker1992}; cf. also the discussion after Eq.~(\ref{eq_admittance}).}. For a good $S$-$S'$ contact, $\widetilde{g}_T$ can be many orders of magnitudes higher than $g_T$, such that $\widetilde{E}_J\gg E_J$. We will later discuss how this classical picture can be justified from a full quantum mechanical model (for details, see also Appendix~\ref{app:quantum_circuit_dissipation}). 

From Eq.~(\ref{eq:phivarphi}) we find
the condition for the resonance frequencies
\begin{equation}
\begin{split}
\alpha^{2}\omega^{4}=\left[\omega^{2}-2i\omega\frac{Y\left(\omega\right)}{C+C_0}-\omega_{0}^{2}\left(1-\alpha^{2}\right)\right]\\\times\left[\omega^{2}-2i\omega\frac{\widetilde{Y}\left(\omega\right)}{\widetilde{C}+C_0}-\widetilde{\omega}_{0}^{2}\left(1-\alpha^{2}\right)\right].\label{eq:eigenmodes_circuit}
\end{split}
\end{equation}
Here, the frequencies $\omega_{0}=\sqrt{8E_{C}E_{J}}$ and $\widetilde{\omega}_{0} =\sqrt{8\widetilde{E}_{C}\widetilde{E}_J}$ reduce to the resonance frequencies of the two $LC$ circuits in the absence of coupling and dissipation.
The charging energies are
\begin{align}
E_{C}&=\frac{e^2}{2}\frac{\widetilde{C}+C_{0}}{C\widetilde{C}+\left[C+\widetilde{C}\right]C_{0}}\\
\widetilde{E}_{C}&=\frac{e^2}{2}\frac{C+C_{0}}{C\widetilde{C}+\left[C+\widetilde{C}\right]C_{0}}\ ,
\end{align}
and the dimensionless coupling parameter
is 
\begin{equation}\label{eq_alpha_def}
\alpha=\frac{C_{0}}{\sqrt{\left(C+C_{0}\right)\left(\widetilde{C}+C_{0}\right)}}\ .
\end{equation}
We then make two main assumptions. We assume both a small coupling between the two resonators, and a small dissipation (i.e., transmons with a high quality factor). In this limit, we expand $\omega$ around the transmon's resonance frequency, $\omega\approx\omega_{0}+\delta\omega$, with a small correction $\delta\omega\ll \omega_0$.

This correction has both a real and an imaginary part. The real part corresponds to merely a small shift of the transmons resonance frequency, owing to the coupling. It amounts to
\begin{equation}
\text{Re}\,\delta \omega=-\frac{1}{2}\frac{\alpha^2\widetilde{\omega}_0^2}{\widetilde{\omega}_0^2-\omega_0^2}\omega_0\ .
\end{equation}
Assuming $\widetilde{\omega}_0>\omega_0$, we find that the necessary condition to justify the perturbative expansion is a sufficiently large detuning $\widetilde{\omega}_0^2-\omega_0^2\gg \alpha^2 \widetilde{\omega}_0^2$, which is justified for realistic parameters (see below). Once this condition is satisfied, it is easy to see that $\delta\omega\ll\omega_0$. In fact, we will show that for our system $\widetilde{\omega}_0\gg\omega_0$, where above condition is equivalent to $\alpha\ll 1$.

Under the same assumptions, we receive for the imaginary part of the correction
\begin{equation}\label{eq_RC_times}
\text{Im}\,\delta\omega=\tau_{RC}^{-1}+\widetilde{\tau}_{RC}^{-1}\ .
\end{equation}
The first term in Eq.~\eqref{eq_RC_times}
corresponds to the $RC$-time of the transmon itself,
\begin{equation}\label{eq_RCtime_0}
\tau_{RC}^{-1}=\frac{\text{Re}\left[Y\left(\omega_{0}\right)\right]}{C+C_0}\ ,
\end{equation}
where the losses
arise from quasiparticles tunneling across the junction~\cite{prb1}. The second term is the sought after correction,
\begin{equation}\label{eq_RCtime_corr}
\widetilde{\tau}_{RC}^{-1}=\frac{\alpha^{2}\omega_{0}^{4}}{\left(\widetilde{\omega}_{0}^{2}-\omega_{0}^{2}\right)^{2}}\frac{\text{Re}\left[\widetilde{Y}\left(\omega_{0}\right)\right]}{\widetilde{C}+C_0}\ ,
\end{equation}
due to quasiparticles
tunneling across the $S$-$S'$ interface.

In order to proceed, we first need an expression
for the admittances which is valid 
for both low and high contact transparency,
in order to account for the improved $S$-$S'$ interface. We assume for both $Y$ and $\widetilde{Y}$ a junction with the same gap parameter on either side. 
This means that for the $S$-$S'$ junction, we need to assume a sufficiently strong proximity effect, such that the
 gap is equal to $\widetilde{\Delta}$ on both sides. In this case, we can use the general formula
provided by Ref.~\cite{Kos2013}
\begin{equation}
\begin{split}
\text{Re}\left[Y\left(\omega\right)\right]=\sum_{n}\frac{1}{\omega}\frac{e^{2}\tau_{n}}{\pi}\int_{\Delta}^{\infty}\!d\epsilon\,\nu\left(\epsilon\right)\nu\left(\omega+\epsilon\right)\\\times\left|z_{n}\right|^{2}\left[f\left(\epsilon\right)-f\left(\omega+\epsilon\right)\right]
\end{split}
\end{equation}
with
\begin{equation}
    \nu(\epsilon) = \frac{\epsilon\sqrt{\epsilon^2-\Delta^2}}{\epsilon^2 - \Delta^2\left[1-\tau_{n}\sin^{2}\left(\frac{\varphi}{2}\right)\right]}
\end{equation}
and
\begin{equation}
\left|z_{n}\right|^{2}=\left|1+\frac{\Delta^{2}\left[1+\cos\varphi\right]-\Delta^{2}\left[1-\tau_{n}\sin^{2}\left(\frac{\varphi}{2}\right)\right]}{\epsilon\left(\omega+\epsilon\right)}\right|^{2}\ .
\end{equation}
The two distribution functions $f(\epsilon)$ and $f(\epsilon+\omega)$ represent respectively the absorption and emission of an energy quantum by the quasiparticle reservoir.
In order to obtain $\widetilde{Y}$ we simply take above expressions and replace $\varphi\rightarrow\widetilde{\varphi}$, $\Delta\rightarrow\widetilde{\Delta}$, $\tau_n\rightarrow\widetilde{\tau}_n$ (the latter standing for the channel transparencies of the $S$-$S'$ junction), and $f\rightarrow\widetilde{f}$ (likewise denoting the quasiparticle distribution function at the $S$-$S'$ junction).
Note that the authors in~\cite{Kos2013} use a different definition of the phase difference $\varphi$ than we do. We account for this, by dividing their original result by a factor 2.

In the limit $E_C\ll E_J$ and $\widetilde{E}_C\ll\widetilde{E}_J$, we have both
$\varphi\approx0$ and $\widetilde{\varphi}\approx0$, in accordance with the above approximation
of a harmonic oscillator close to ground state, and we find
\begin{equation}
\begin{split}
& \text{Re}\left[\widetilde{Y}\left(\omega\right)\right]\approx\sum_{n}\frac{1}{\omega}\frac{e^{2}\widetilde{\tau}_{n}}{\pi}\int_{\widetilde{\Delta}}^{\infty}d\epsilon
\frac{\epsilon}{\sqrt{\epsilon^2-\widetilde{\Delta}^2}}
\\&\times
\frac{\epsilon+\omega}{\sqrt{(\epsilon+\omega)^2-\widetilde{\Delta}^2}}
\left|1+\frac{\widetilde{\Delta}^{2}}{\epsilon\left(\omega+\epsilon\right)}\right|^{2}\left[f\left(\epsilon\right)-f\left(\omega+\epsilon\right)\right].
\end{split}
\end{equation}
Let us now further assume that the quasiparticle distribution functions have a finite effective temperature $T_\text{qp}$, such that $f$ ($\widetilde{f}$) only has a finite support for $\epsilon$ between $\Delta$ ($\widetilde{\Delta}$) and $\Delta+T_\text{qp}$ ($\widetilde{\Delta}+T_\text{qp}$). In particular, the validity of this assumption for $\widetilde{f}$ has to be discussed with care. We expect it to be valid for two reasons. First of all, for the desired regime of a good contact and a large gap difference, the relaxation rate of quasiparticles at energies $\sim\Delta$ approaches the normal metal limit [see Eq.~\eqref{eq_tau_th}] and is therefore fast compared to the qubit relaxation time~\cite{Riwar}. Secondly, even if there remains some occupation of quasiparticles at energies close to $\Delta$, it is proportional to the steady-state occupation at $\epsilon\approx\Delta$, which will be 
smaller than the steady-state occupation at $\widetilde{\Delta}$; the latter relaxes only via pair annihilation processes, which provide the bottleneck time scale at low temperatures~\cite{Kaplan1976}. Therefore, for the integral over all energies, such a residual occupation at high energies will provide but a small correction to $\widetilde{Y}$. In any case, we find
\begin{equation}\label{eq_admittance}
\text{Re}\left[Y\left(\omega\right)\right]=\frac{2}{\omega}\Delta g_{T}\int_{\Delta}^{\infty}d\epsilon\frac{f\left(\epsilon\right)-f\left(\omega+\epsilon\right)}{\sqrt{\epsilon-\Delta}\sqrt{\epsilon+\omega-\Delta}}\ ,
\end{equation}
having in addition defined the normal metal conductance as $g_{T}=2 g_0 \sum_{n}\tau_{n}$. The expression for $\widetilde{Y}$ follows again in analogy, with the additional replacement $g_T\rightarrow\widetilde{g}_T$, where $\widetilde{g}_{T}=2g_0 \sum_{n}\widetilde{\tau}_{n}$.
In fact, we have arrived at the same expression as in Ref.~\cite{prb1},
where it was derived for tunneling SIS junctions. We thus verified that in the nearly harmonic regime we are considering, this expression is valid for channels with both high and low transparency.

To continue, we need to discuss the dependence of the admittance with respect to the quasiparticle temperature. We can consider both a regime of ``cold'' quasiparticles, $\omega\gg T_\text{qp}$, and ``hot'' quasiparticles $\omega\ll T_\text{qp}$. The former regime is simpler, which is why we discuss it first. Here, the emission of energy quanta by the quasiparticle reservoir is suppressed, and we find
\begin{equation}\label{eq_ReY_cold}
\text{Re}\left[Y\left(\omega\right)\right] \approx\frac{1}{2}\left(\frac{2\Delta}{\omega}\right)^{\frac{3}{2}}g_{T}x_{\text{qp}}\ ,
\end{equation}
which is in accordance with the expression found in~\cite{prb1}. In fact, with $Y$ and $\widetilde{Y}$ in the limit of cold quasiparticles, it is possible to show that the $RC$ times in Eqs.~\eqref{eq_RCtime_0} and~\eqref{eq_RCtime_corr} follow directly from the full quantum mechanical calculation, shown in Appendix~\ref{app:quantum_circuit_dissipation}.

Recent measurements of transmon transition rates~\cite{Serniak_2018} have been interpreted in terms of hot non-equilibrium quasiparticles, although a more likely explanation is in terms of photon-assisted Cooper pair breaking~\cite{Houzet2019}. In any case, if there are hot quasiparticles (in or out of equilibrium), the emission process can no longer be ignored. What is more, the classical picture breaks down, a fact we have already discussed in the different context of a model for normal-metal traps~\cite{Riwar2018}. Hence, the circuit relaxation time can no longer be strictly considered an $RC$ time. One can nonetheless restore an effectively classical model with a modified admittance, $Y_\text{mod}$, where the absorption and emission processes are \textit{added} instead of subtracted, as in Eq.~\eqref{eq_admittance}. One can justify this procedure by noting that in the actual quantum mechanical model there occur the rates of qubit relaxation $\Gamma_{1\rightarrow 0}$ and excitation $\Gamma_{0\rightarrow 1}$. The relaxation rate of the qubit is then given by the \textit{sum}, $\tau^{-1}=\Gamma_{1\rightarrow 0}+\Gamma_{0\rightarrow 1}$. We provide more details on this argument in Appendix~\ref{app:quantum_circuit_dissipation}. For the modified admittance, with $\omega\ll T_\text{qp}$, we find,
\begin{equation}\label{eq_ReY_hot}
\text{Re}\left[Y_\text{mod}(\omega)\right]\approx \sqrt{\frac{\omega}{4T_{\text{qp}}}}\ln\left(\frac{4T_{\text{qp}}}{\omega}\right)\left(\frac{2\Delta}{\omega}\right)^{\frac{3}{2}}g_{T}x_{\text{qp}}\ .
\end{equation}
We see that this result differs from the cold quasiparticle result, Eq.~\eqref{eq_ReY_cold}, merely by a prefactor $\ln(r)/\sqrt{r}$, depending on the ratio $r=4T_\text{qp}/\omega_0$, which contains no parameters specific to the $S$-$S'$ junction, i.e., it is the same for both $Y$ and $\widetilde{Y}$. This will simplify the discussion at the end of this section. 

Finally, we need to provide estimates for the capacitances. First of all, both $C$ and
$C_{0}$ are coplanar capacitors, with different distances between
the two plates, $D$ for $C$ and $D+a$ for $C_{0}$, see also Fig.~\ref{fig_SSp_model}a. Consequently,
we can approximate their ratio by means of the electrostatic result for coplanar capacitors (see Appendix~\ref{app:coplanar}). In order to determine $C$, we compute the relation between the accumulated charge across the transmon junction with respect to an applied voltage difference $V$, such that
\begin{equation}
C\approx \frac{\epsilon_0}{\pi}W\ln\left(\frac{4L}{D}\right)\ .
\end{equation}
valid for $D\ll L$. With $W$ we denote the width of the superconducting plates and with $\epsilon_0$ the vacuum permittivity. Note that due to the logarithmic dependence on $L/D$, this geometric factor contributes but a numerical prefactor, which we can discard for our estimate, $C\sim\epsilon W$. As for the capacitance $C_0$, we may in principle take into account that the trap part can have a varying length $L-a$; it can be either close to $L$, such that $a\lesssim D$ and $D+a\ll L-a$, or we could also consider a small trap $D+a\gg L-a$. However both scenarios provide again only a logarithm of geometric factors, such that likewise $C_0\sim\epsilon W$. 
We  observe, that while $C$ 
is generally larger
than $C_0$, 
the two are realistically within one order of magnitude of each other (see Appendix~\ref{app:coplanar}). We therefore use $C\sim C_{0}$ in the following.

Concerning the capacitance of the
second oscillator $\widetilde{C}$, we treat it likewise as a geometric capacitance. However, unlike the previous two capacitances, we here assume that the $S'$ layer is grown on top of the $S$ layer, such that we have an ordinary parallel plate capacitor,
\begin{equation}
\widetilde{C}=\frac{\epsilon_0 \widetilde{A}}{d}
\end{equation}
with $\widetilde{A}=W(L-a)$ being the contact area, and $d$ the thickness of the barrier between the two layers.
(see Fig.~\ref{fig_SSp_model}). 
Because of the good $S$-$S'$ contact, $d$ is very small, $L-a\gg d$, such that we find
$\widetilde{C}\gg C,\,C_{0}$, irrespective of the geometric details.
As a consequence, we have
$\alpha\sim\sqrt{C/\widetilde{C}} \ll 1$,
thanks to the dominant $\widetilde{C}$. This validates our initial
hypothesis of the circuits being weakly coupled, in spite of $C\sim C_{0}$.

Having expressions for the capacitances at our hands, we now proceed by comparing the resonance frequencies of the two oscillators. Importantly, both $\widetilde{C}\gg C$
and $\widetilde{E}_J\gg E_{J}$, such that it is not yet obvious whether $\widetilde{\omega}_0$ is much larger than $\omega_0$ or not. As for the first factor, we expect $d$ to be at most on the nanometer scale, whereas $L-a$ can be as large as hundreds of micrometers to a millimeter, resulting in $\widetilde{C}/C\lesssim10^6$. As for the Josephson energies, we estimate their ratio as follows.
We use the relation between the Josephson energies and their respective contact conductances, given above. We therefore need to estimate the ratio $g_T/\widetilde{g}_{T}$. In order to do so, let us assume for simplicity that all channels have roughly the same transparency, such that $g_T/g_0\approx N_\text{ch}\langle\tau\rangle$, where $N_\text{ch}$ is the number of channels, and $\langle\tau\rangle$ is the average over the channel ensemble of the transmission probabilities (and similarly for $\widetilde{g}_T$). For weakly-coupled tunnel junctions, we take $\langle\tau\rangle\sim 10^{-5}$~\cite{Riwar}; this estimates applies to $g_T$. For very good contacts (i.e., for $\widetilde{g}_T$) on the other hand, $\langle\tau\rangle$ should in principle be much larger, of the order $10^{-1}$ to $1$; this is needed to ensure strong proximity effect, and it is confirmed by experiments~\cite{Aarts1997,Kyle}. In order to estimate the ratio of channel numbers, we note that the number of channels scales with the contact area, such that $N_\text{ch}/\widetilde{N}_\text{ch}=A/\widetilde{A}$, where $A$ is the contact area of the transmon junction. This ratio depends strongly on the transmon and trap geometry. Typically, $A\sim (0.1\,\mu\text{m})^2$. For a small trap, $\widetilde{A}$ could be as low as $10\,\mu\text{m}\times 100\,\mu\text{m}$. A large trap, covering a good portion of the transmon, can be at least one order of magnitude larger ($\sim 100\,\mu\text{m}\times 100\,\mu\text{m}$). Let us take the small trap and $\langle\widetilde{\tau}\rangle\sim 10^{-1}$ as an upper bound for the ratio, such that
\begin{equation}\label{eq_gT_ratio}
\frac{g_T}{\widetilde{g}_T}<10^{-9}\ .
\end{equation}
The ratio $E_J/\widetilde{E}_J$ between the Josephson energies may have a slightly less stringent upper bound, due to the additional factor $\Delta/\widetilde{\Delta}>1$; however, the two energy gaps will differ in practice by at most one order of magnitude, since we require $\omega_0<2\widetilde{\Delta}$ to avoid pair breaking processes. Consequently, we find that
\begin{equation}
\frac{\widetilde{\omega}_0}{\omega_0}>10
\end{equation}
indicating that the increase of the contact conductance is the dominating effect. This means that the first excited state of the transmon remains the lowest energetically accessible state, such that we do not have to concern ourselves with any relaxation process of the qubit through exciting the $SS'$ oscillator.

Finally, we have all elements at disposal to estimate the $RC$ times. We first discuss the contribution of the $S$-$S'$ junction, in particular its dependence on the junction conductance. Due to $\widetilde{\omega}_0\gg \omega_0$ and $\widetilde{C}\gg C\sim C_0$, we find
\begin{equation}
\widetilde{\tau}_{RC}^{-1} \sim \frac{1}{C} \sqrt{\frac{\Delta}{\widetilde{\Delta}}}\frac{g_{T}}{\widetilde{g}_{T}}\left(\frac{\Delta}{\omega_{0}}\right)^{\frac{3}{2}}g_{T}\widetilde{x}_{\text{qp}}\ .
\end{equation}
We note that the above result is correct as such for cold quasiparticles, whereas for the hot regime we would have to include the prefactor depending on $T_\text{qp}/\omega_0$, as identified in Eq.~\eqref{eq_ReY_hot}. We discard it for lack of relevance: as mentioned right after Eq.~\eqref{eq_ReY_hot}, this prefactor does not depend on junction specific parameters (that is, it is independent of $\widetilde{g}_T$).
With the above result, we can answer the first question whether an increase in the interface quality increases or decreases the qubit relaxation rate. As we see, even though the admittance increases linearly with $\widetilde{g}_T$, the prefactor $1/\widetilde{\omega}_0^4\sim 1/\widetilde{g}_T^2$ decreases faster than that; hence the positive effect of the decoupling due to frequency mismatch dominates over the negative effect of the increased tunneling rate. 
Improving the $S$-$S'$ interface is therefore beneficial.

We now compare the two $RC$ times, in order to ensure that $\widetilde{\tau}_{RC}>\tau_{RC}$. As mentioned above, otherwise we would encounter a counterproductive case where the trap dissipation due to quasiparticle tunneling would be larger than the original quasiparticle process the trap was designed to mitigate.
The $RC$-time of the transmon can likewise be expressed in terms of circuit parameters and quasiparticle density as (for cold quasiparticles)
\begin{align}
\tau_{RC}^{-1}
 & \sim\frac{1}{C}\left(\frac{\Delta}{\omega_{0}}\right)^{\frac{3}{2}}g_{T}x_{\text{qp}}.
\end{align}
Through this, we find,
\begin{equation}
\frac{\tau_{RC}}{\widetilde{\tau}_{RC}}\sim\sqrt{\frac{\Delta}{\widetilde{\Delta}}}\frac{g_{T}}{\widetilde{g}_{T}}\frac{\widetilde{x}_{\text{qp}}}{x_{\text{qp}}}\ .
\end{equation}
This result is valid for both cold and hot quasiparticles [as the prefactor identified in Eq.~\eqref{eq_ReY_hot} drops out from the ratio]. Importantly, the ratio between the two $RC$ times has both large and small factors. While the ratio of gaps satisfies $\sqrt{\Delta/\widetilde{\Delta}}>1$, it will not be a significant factor, and can be simplified to $\sim1$. The improved contact conductance works in our favour, and provides a factor of at least $10^{-9}$, see Eq.~\eqref{eq_gT_ratio}.
As far as the quasiparticle density is concerned, it has been shown~\cite{trapopt} that a good trap can reduce $x_\text{qp}$ at the junction by about two orders of magnitude with respect to the stationary density $x_\text{qp}^\text{st}$ in the absence of the trap. That is, for a typical value $x_\text{qp}^\text{st}\sim 10^{-6}$ we can expect ideally $x_\text{qp}\sim 10^{-8}$ at the junction. The quasiparticle density at the $S$-$S'$ junction $\widetilde{x}_\text{qp}$ on the other hand can be significantly higher than $x_\text{qp}^\text{st}$. After all, the trap has a lower gap, $\widetilde{\Delta}<\Delta$, and is designed to attract and collect nonequilibrium quasiparticles from all over the device, which recombine very slowly. However, in order to reach the regime $\tau_{RC}^{-1}\sim\widetilde{\tau}_{RC}^{-1}$, $\widetilde{x}_\text{qp}$ would have to be larger than unity, implying the full suppression of superconductivity in the trap. We expect this to be very far from a realistic scenario for transmon qubits: as we have already noted, even in experiments with quasiparticle injection, the nonequilibrium quasiparticle density accumulated at the junction does not exceed $10^{-3}$~\cite{wang,Riwar}. We can therefore safely conclude that $\widetilde{\tau}_{RC}\gg\tau_{RC}$. Consequently, the dissipation due to quasiparticle tunneling across the $S$-$S'$ junction cannot severely impede the qubit quality factor, in spite of the increased contact conductance and the high quasiparticle density in the trap.

\section{Summary}\label{sec_conclusions}

We have theoretically examined the quasiparticle dynamics and dissipative processes of gap-engineered traps in transmon qubits. In our model, the gap engineering is realized through the proximity effect, by coupling a second, lower gap superconductor to the transmon, with a good contact of high transparency.

We have identified several advantages with respect to normal-metal traps studied in earlier works. As far as the quasiparticle diffusion is concerned we have predicted that gap-engineered traps, unlike normal-metal traps, do not suffer from quasiparticle backflow (i.e., escape from the trap back to the qubit). As a consequence, the minimal trap size above which the traps are in a regime of strong and efficient quasiparticle evacuation is reduced. Therefore, strong gap-engineered traps are available for even smaller devices. Concerning dissipative processes, we have studied the ac resistivity of the bulk superconductor, and found that the resistivity drops by many orders of magnitudes with respect to the normal metal case. Consequently, bulk dissipation arising due to charge redistribution, which was a critical issue for normal metal traps, is no longer of any concern.

Finally, we have investigated losses due to quasiparticle tunneling at the proximity interface. We have found that even though the admittance of the interface increases significantly, due to the combination of a good contact and an increased quasiparticle density in the trap region, the overall influence of the trap is negligible. The main reason for this is that a good contact strongly detunes the transmon and trap resonance frequencies, and effectively decouples the two. Given the above advantages, we come to the conclusion that gap engineering may be the preferable strategy to implement quasiparticle traps.

\acknowledgments

We gratefully acknowledge useful discussions with A. Hosseinkhani and K. Serniak. This work was supported in
part by a Max Planck award (R.-P.~R.).

\appendix

\section{Justification of diffusion equation from microscopic equations}
\label{app:diffusion}

In the main text, we provide heuristic boundary conditions for the quasiparticle density $x_\text{qp}$ at the the trap edge, where the order parameter drops. We here show how to justify these boundary conditions from a microscopic vantage point.

For a proximitized piece of a dirty superconductor with an inhomogeneous gap parameter $\Delta(x)$, the diffusion equation providing the dynamics of the quasiparticle occupation number as a function of energy $f_{\text{qp}}(\epsilon)$ can be given as~\cite{Belzig1999May}
\begin{equation}\label{diffusion_of_E}
\dot{f}_{\text{qp}}\left(\epsilon,y\right)=\partial_{y}\left[D_{\text{qp}}\left(\epsilon,y\right)\partial_{y}f_{\text{qp}}\left(\epsilon,y\right)\right]\ ,
\end{equation}
where we assume that the charge imbalance contribution is negligible. In the above equation, the inhomogeneous gap $\Delta\left(y\right)$ enters through the $y$-dependent diffusion coefficient $D_{\text{qp}}\left(\epsilon,y\right)=D_{0}/\nu_\text{BCS}(\epsilon)$, where $D_{0}$ is the normal state diffusion coefficient, and $\nu_\text{BCS}(\epsilon)=\epsilon/\sqrt{\epsilon^{2}-\Delta^{2}\left(y\right)}$ is the normalized BCS density of states.

We know that the gap parameter for a proximitized slab of superconductor varies on the length scale given by the superconducting coherence length $\xi$ (see, e.g., Ref.~\cite{Cherkez2014}), which is much smaller than the length scales relevant for the diffusion of quasiparticles. We can therefore separate above equation into two parts with a constant gap, a first part for $0<y<a$ with gap $\Delta(y)=\Delta$ and a second part for $a<y<L$ with $\Delta(y)=\widetilde{\Delta}$ (we assume $a,\, L-a \gg \xi$). The region where the gap abruptly jumps can now be incorporated through the boundary conditions [which are straightforwardly derived from Eq.~\eqref{diffusion_of_E}],
\begin{align}
f^-_\text{qp}(\epsilon)&=f^+_\text{qp}(\epsilon)\label{eq_bc_fqp}\\
D^-_\text{qp}(\epsilon)\partial_y f^-_\text{qp}(\epsilon)&=D^+_\text{qp}(\epsilon)\partial_y f^+_\text{qp}(\epsilon)\label{eq_bc_dfqp}
\end{align}
with the notation $q^-=\left.q\right|_{y=a-0^+}$ and $q^+=\left.q\right|_{y=a+0^+}$. We stress that in Eq.~\eqref{eq_bc_dfqp}, the diffusion constant does not cancel, because it is not the same for $y=a-0^+$ and $y=a+0^+$.

We note that the derivation of an effective boundary for $x_\text{qp}$ based on above Eqs.~\eqref{eq_bc_fqp} and~\eqref{eq_bc_dfqp} is not trivial, due to the density of states being position dependent. We can however use a certain set of assumptions on the function $f_\text{qp}$ through which we can still find an effective boundary condition for $x_\text{qp}$. The procedure goes as follows.

First, we assume that the quasiparticle distribution function can be written as a product, $f_\text{qp}(\epsilon,y)=g(\Delta \epsilon)*h(y)$, where $\Delta \epsilon=\epsilon-\Delta(y)$. We furthermore assume that $g(\Delta \epsilon)$ has a finite support within the window $0<\Delta \epsilon<k_\text{B}T_\text{qp}$, which we refer to as the quasiparticle temperature $T_\text{qp}$. We focus on the regime of cold quasiparticles, $k_\text{B}T_\text{qp}\ll \Delta(y)$ (we note that this set of approximations has already been proven useful when describing experiments, see e.g., \cite{Riwar}). For the free and the proximitized parts we get independently the diffusion equations on either side of $y=a$ as,
\begin{equation}
\dot{x}_\text{qp}(y)=D_\text{qp}^{\mp}\partial_y^2 x_\text{qp}(y)\ ,
\end{equation}
where the phenomenological diffusion coefficient is given as
\begin{equation}
D_\text{qp}^\mp=\frac{\int_{\Delta^{\mp}}^\infty d\epsilon \nu_\text{BCS}(\epsilon)g(\epsilon-\Delta^{\mp}) D_\text{qp}(\epsilon)}{\int_{\Delta^{\mp}}^\infty d\epsilon \nu_\text{BCS}(\epsilon)g(\epsilon-\Delta^{\mp}) }\ ,
\end{equation}
with $\Delta^-=\Delta$ and $\Delta^{+}=\widetilde{\Delta}$, for $y<a$ and $y>a$, respectively. Under the above assumptions, we indeed find $D_\text{qp}^- \ll D_\text{qp}^+$. We can show this explicitly, by making the crude simplification, $g(\Delta \epsilon)\approx \theta(\Delta \epsilon)\theta(k_\text{B}T_\text{qp}-\Delta \epsilon)$, such that $D_\text{qp}^{\mp}\sim \sqrt{k_\text{B}T_\text{qp}/\Delta^{\mp}}D_0$. We stress however, that the above result holds qualitatively (up to numerical prefactors) also for more realistic assumptions for $g$, as long as $g$ does not have a too pronounced energy dependence (apart from the cutoff).

We now derive explicitly the boundary conditions. We choose to multiply Eqs.~\eqref{eq_bc_fqp} and~\eqref{eq_bc_dfqp} with the BCS density of states on the non-proximitized side, where $\Delta(y)=\Delta$, because $x_\text{qp}$ at the junction is of most interest to us. Under the same assumptions as above, integrating the resulting equations provides us with the following expressions:
\begin{align}
\label{eq_bc_xqp_app}x_{\text{qp}}^{-}&=\sqrt{2}\sqrt{\frac{\Delta^{2}-\widetilde{\Delta}^{2}}{\Delta T_{\text{qp}}}}x_{\text{qp}}^{+}\ ,\\
\label{eq_bc_dxqp_app}D_{\text{qp}}^{-}\partial_{y}x_{\text{qp}}^{-}&=\sqrt{2}\sqrt{\frac{\Delta^{2}-\widetilde{\Delta}^{2}}{\Delta T_{\text{qp}}}}D_{\text{qp}}^{+}\partial_{y}x_{\text{qp}}^{+}\ .
\end{align}
To arrive at above equations, we have furthermore assumed that $T_\text{qp}\ll\Delta-\widetilde{\Delta}$. Importantly, both Eq.~\eqref{eq_bc_xqp_app} and~\eqref{eq_bc_dxqp_app} contain the same correction prefactor. We can thus conclude that the heuristic boundary condition given in the main text, Eq.~\eqref{eq_bc_xqp}, can be justified from the microscopic starting point here, Eq.~\eqref{diffusion_of_E}, when renormalizing $x_\text{qp}$ for $y>a$ by this prefactor.

\section{Dissipation due to photo-assisted tunneling in the quantum circuit approach}
\label{app:quantum_circuit_dissipation}

In the main text, we deploy a classical model for the dynamics of the transmon with an $SS'$ trap, as coupled $LC$ circuits. Here, we indicate how one can justify the classical model by means of a full quantum mechanical model. 

For this purpose, we first compute the non-dissipative part of the dynamics, disregarding the effects of the quasiparticles.
Our starting point is the Lagrangian for the phase differences  $\varphi$ and  $\widetilde\varphi$ across the transmon and $S$-$S'$ junction, respectively, 
\begin{align}
L&=\frac{1}{2}\frac{C}{\left(2e\right)^{2}}\dot{\varphi}^{2}+E_{J}\cos\left(\varphi\right)\nonumber\\&+\frac{1}{2}\frac{\widetilde{C}}{\left(2e\right)^{2}}\dot{\widetilde{\varphi}}^{2}+\sum_{n}\widetilde{\Delta}\sqrt{1-\tau_{n}\sin^{2}\left(\frac{\widetilde{\varphi}}{2}\right)}\nonumber\\&+\frac{1}{2}\frac{C_{0}}{\left(2e\right)^{2}}\left(\dot{\varphi}-\dot{\widetilde{\varphi}}\right)^{2}\ ,
\end{align}
with the same capacitances as in the model given in Fig.~\ref{fig_dissipative_circuit}. We perform the Legendre transformation to arrive at the Hamiltonian.  The variables conjugate to the phase difference operators are
\begin{align}
n&=\frac{\partial L}{\partial\dot{\varphi}}=\frac{C+C_{0}}{\left(2e\right)^{2}}\dot{\varphi}-\frac{C_{0}}{\left(2e\right)^{2}}\dot{\widetilde{\varphi}}\ ,\\\widetilde{n}&=\frac{\partial L}{\partial\dot{\widetilde{\varphi}}}=\frac{\widetilde{C}+C_{0}}{\left(2e\right)^{2}}\dot{\widetilde{\varphi}}-\frac{C_{0}}{\left(2e\right)^{2}}\dot{\varphi}\ ,
\end{align}
with the quantization conditions $[n,\varphi]=i$ and $[\widetilde{n},\widetilde{\varphi}]=i$, respectively.
Furthermore, we assume small phase differences $\varphi\approx0$ and $\widetilde{\varphi}\approx 0$. Thus we arrive at a system of two coupled harmonic oscillators,
\begin{equation}
\begin{split}
H=\frac{1}{2}E_C n^{2}+\frac{1}{2}E_{J}\varphi^{2}+\frac{1}{2}\widetilde{E}_{C}\widetilde{n}^{2}\\+\frac{1}{2}\widetilde{E}_{J}\widetilde{\varphi}^{2}+E_{C_0}n\widetilde{n}\ ,
\end{split}
\end{equation}
with
\begin{align}
E_{C}&=\left(2e\right)^{2}\frac{\widetilde{C}+C_{0}}{C\widetilde{C}+\left[C+\widetilde{C}\right]C_{0}}\\\widetilde{E}_{C}&=\left(2e\right)^{2}\frac{C+C_{0}}{C\widetilde{C}+\left[C+\widetilde{C}\right]C_{0}}\\E_{C_{0}}&=\left(2e\right)^{2}\frac{C_{0}}{C\widetilde{C}+\left[C+\widetilde{C}\right]C_{0}}\ .
\end{align}
Here to simplify the comparison with textbook formulas for the harmonic oscillator, we use a different definition of charging energies compared to the standard one adopted in the main text.
With these definitions, we see that the harmonic oscillator approximation is valid for $E_C\ll E_{J}$ as well as $\widetilde{E}_{C}\ll \widetilde{E}_J$ (we remind, $\widetilde{E}_J=\widetilde{\Delta}/4\sum_n\tau_n$). We continue by introducing ladder operators for the harmonic oscillators,
\begin{align}
n&=\frac{i}{\sqrt{2}}\left(\frac{E_J}{E_C}\right)^{1/4}\left(a^{\dagger}-a\right)\ ,\\ \varphi&=\frac{1}{\sqrt{2}}\left(\frac{E_C}{E_J}\right)^{1/4}\left(a^{\dagger}+a\right)\ .
\end{align}
The operators $\widetilde{n}$ and $\widetilde{\varphi}$ are obtained in analogy, by replacing $E_C\rightarrow \widetilde{E}_{C}$ and $E_J\rightarrow \widetilde{E}_J$. We arrive at
\begin{align}
H&=\omega_{0}\left(a^{\dagger}a+\frac{1}{2}\right)+\widetilde{\omega}_{0}\left(\widetilde{a}^{\dagger}\widetilde{a}+\frac{1}{2}\right)\nonumber\\&-\frac{1}{2}\alpha\sqrt{\omega_{0}\widetilde{\omega}_{0}}\left(a^{\dagger}-a\right)\left(\widetilde{a}^{\dagger}-\widetilde{a}\right)\ .
\end{align}
with
\begin{equation}
\alpha=\frac{E_{C_{0}}}{\sqrt{E_{C}\widetilde{E}_{C}}}=\frac{C_{0}}{\sqrt{\left(C+C_{0}\right)\left(\widetilde{C}+C_{0}\right)}}\ ,
\end{equation}
just as in the classical equation in the main text, Eq.~\eqref{eq_alpha_def}.
The eigenstates for zero coupling can be denoted as $|m,\widetilde{m}\rangle=(a^\dagger)^m({\widetilde{a}}^\dagger)^{\widetilde{m}}|0,0\rangle$, where $|0,0\rangle$ is the ground state without excitations.

We now perform a perturbation theory for a small coupling parameter, $\alpha\ll 1$. In lowest order, we receive the corrections for the general circuit states
\begin{equation}
\left|m,\widetilde{m}\right\rangle \rightarrow \left|m,\widetilde{m}\right\rangle +\left|\delta_{m,\widetilde{m}}\right\rangle .
\end{equation}
The corrections have the following nonzero elements
\begin{align}
\left\langle m+1,\widetilde{m}+1|\delta_{m,\widetilde{m}}\right\rangle &=\frac{\alpha}{2}\sqrt{\omega_{0}\widetilde{\omega}_{0}}\frac{\sqrt{\left(m+1\right)\left(\widetilde{m}+1\right)}}{\widetilde{\omega}_{0}+\omega_{0}}\\\left\langle m+1,\widetilde{m}-1|\delta_{m,\widetilde{m}}\right\rangle &=\frac{\alpha}{2}\sqrt{\omega_{0}\widetilde{\omega}_{0}}\frac{\sqrt{\left(m+1\right)\widetilde{m}}}{\widetilde{\omega}_{0}-\omega_{0}}\\\left\langle m-1,\widetilde{m}+1|\delta_{m,\widetilde{m}}\right\rangle &=-\frac{\alpha}{2}\sqrt{\omega_{0}\widetilde{\omega}_{0}}\frac{\sqrt{m\left(\widetilde{m}+1\right)}}{\widetilde{\omega}_{0}-\omega_{0}}\\\left\langle m-1,\widetilde{m}-1|\delta_{m,\widetilde{m}}\right\rangle &=-\frac{\alpha}{2}\sqrt{\omega_{0}\widetilde{\omega}_{0}}\frac{\sqrt{m\widetilde{m}}}{\widetilde{\omega}_{0}+\omega_{0}}\ .
\end{align}
We now have all ingredients at hand to compute the dissipation due to quasiparticles.  We are in particular interested in transitions between the qubit states,
\begin{equation}
\left|00\right\rangle, \quad\left|10\right\rangle =a^{\dagger}\left|00\right\rangle 
\end{equation}
assuming that the $SS'$ resonator remains in its ground state. This is a realistic assumption, since $\omega_0\ll\widetilde{\omega}_0$. As shown, e.g., in Ref.~\cite{prb1}, the quasiparticle tunneling process couples to the operators $\sin(\varphi/2)$ (for tunneling across the transmon junction) or $\sin(\widetilde{\varphi}/2)$ (across the $SS'$ junction). In the limit of harmonic dynamics, it suffices to consider the approximation $\sin(\varphi/2)\approx\varphi/2$ (and likewise for $\widetilde{\varphi}$). We then receive
\begin{align}
\left|\left\langle 10\right|\varphi\left|00\right\rangle \right|^{2}&\approx\frac{1}{2}\frac{E_{C}}{\omega_{0}}\\
\left|\left\langle 10\right|\widetilde{\varphi}\left|00\right\rangle \right|^{2}&\approx \frac{1}{2}\frac{\widetilde{E}_{C}}{\omega_{0}}\frac{\alpha^{2}\omega_{0}^{4}}{\left(\widetilde{\omega}_{0}^{2}-\omega_{0}^{2}\right)^{2}}\ ,
\end{align}
up to leading order in the respective transition matrix elements. We note that within the approximation $\alpha\ll1$, the respective charging energies reduce to $E_C\approx(2e)^2/(C+C_0)$ and $\widetilde{E}_C\approx(2e)^2/(\widetilde{C}+C_0)$.
Already on this level we see the appearance of the same prefactor structure as in the classical equation in the main text, Eq.~\eqref{eq_RCtime_corr}.

In order to arrive at the full $RC$ time including the quasiparticle density, we follow Ref.~\cite{prb1}, where it is shown how the above computed transition matrix element enters in the total transition rate. Namely, the transition rates are
\begin{align}
\Gamma_{1\rightarrow0}&=\frac{1}{4}\left|\left\langle 10\right|\varphi\left|00\right\rangle \right|^{2}S_\text{qp}(\omega_0)\\
\widetilde{\Gamma}_{1\rightarrow0}&=\frac{1}{4}\left|\left\langle 10\right|\widetilde{\varphi}\left|00\right\rangle \right|^{2}\widetilde{S}_\text{qp}(\omega_0)\ ,
\end{align}
with
\begin{align}
S_\text{qp}(\omega)&\approx\frac{16E_J}{\pi}\int_\Delta^\infty d\epsilon\frac{f_\text{qp}(\epsilon)}{\sqrt{\epsilon-\Delta}\sqrt{\epsilon+\omega-\Delta}}\\
\widetilde{S}_\text{qp}(\omega)&\approx\frac{16\widetilde{E}_J}{\pi}\int_{\widetilde{\Delta}}^\infty d\epsilon\frac{\widetilde{f}_\text{qp}(\epsilon)}{\sqrt{\epsilon-\widetilde{\Delta}}\sqrt{\epsilon+\omega-\widetilde{\Delta}}}\ .
\end{align}
In the reversed rates $\Gamma_{0\rightarrow 1}$ and $\widetilde{\Gamma}_{0\rightarrow 1}$, there enter the time-reversed correlation functions,
\begin{align}
S_\text{qp}^\text{rev}(\omega)&\approx\frac{16E_J}{\pi}\int_\Delta^\infty d\epsilon\frac{f_\text{qp}(\epsilon+\omega)}{\sqrt{\epsilon-\Delta}\sqrt{\epsilon+\omega-\Delta}}\\
\widetilde{S}^\text{rev}_\text{qp}(\omega)&\approx\frac{16\widetilde{E}_J}{\pi}\int_{\widetilde{\Delta}}^\infty d\epsilon\frac{\widetilde{f}_\text{qp}(\epsilon+\omega)}{\sqrt{\epsilon-\widetilde{\Delta}}\sqrt{\epsilon+\omega-\widetilde{\Delta}}}\ .
\end{align}
Note that we chose $\omega>0$ by default.
The remaining task is to relate the quasiparticle correlation functions to the admittance. They are generally related to each other as
\begin{equation}
S_\text{qp}(\omega)-S_\text{qp}^\text{rev}(\omega)=\frac{2\omega}{\pi}\frac{\text{Re}Y(\omega)}{g_0}\ ,
\end{equation}
and similarly for $\widetilde{Y}$. Let us now again assume that the quasiparticles occupy a finite energy window, given by the effective quasiparticle temperature $T_\text{qp}$. If the quasiparticles are cold with respect to the qubit energy scale, $\omega_0\gg T_\text{qp}$, then we can neglect excitation processes, and the above relation simplifies to the so-called "high-frequency" case,
\begin{equation}
S_\text{qp}(\omega)\approx\frac{2\omega}{\pi}\frac{\text{Re}Y(\omega)}{g_0}\ .
\end{equation}
We finally get
\begin{align}
\Gamma_{1\rightarrow0}&=\frac{1}{4}\frac{E_{C}}{\pi}\frac{\text{Re}Y(\omega_0)}{g_0}\\
\widetilde{\Gamma}_{1\rightarrow0}&=\frac{1}{4}\frac{\widetilde{E}_{C}}{\pi}\frac{\alpha^{2}\omega_{0}^{4}}{\left(\widetilde{\omega}_{0}^{2}-\omega_{0}^{2}\right)^{2}}\frac{\text{Re}\widetilde{Y}(\omega_0)}{g_0}\ ,
\end{align}
which can easily be identified as the respective $RC$ times in the main text, see Eqs.~\eqref{eq_RCtime_0} and~\eqref{eq_RCtime_corr}. 

As already pointed out in the main text, the quasiclassical approximation fails with certainty, if the quasiparticle distribution is hot, $\omega< T_\text{qp}$. Here, we justify the appearance of the modified admittance introduced in the main text, by means of the here described transition rates. We note that we have already commented on a similar scenario in a previous paper~\cite{Riwar2018}. In essence, for hot quasiparticles, both the qubit relaxation and excitation are equally probable, $S_\text{qp}\approx S_\text{qp}^\text{rev}$. The decay rate of the qubit system can now be captured by the "new" $RC$ time, $\tau_{RC}^{-1}=\Gamma_{1\rightarrow0}+\Gamma_{0\rightarrow1}$. This justifies the definition of a modified admittance
\begin{equation}
\text{Re}\left[Y^\text{mod}(\omega)\right]=\frac{2\Delta}{\omega}{g}_{T}\int_{{\Delta}}^{\infty}d\epsilon\frac{f\left(\epsilon\right)+f\left(\omega+\epsilon\right)}{\sqrt{\epsilon-{\Delta}}\sqrt{\epsilon+\omega-{\Delta}}}\ ,
\end{equation}
(and likewise for $\widetilde{\tau}_{RC}$), where the two fermi functions at different energies are added rather than substracted. This equation gives rise to Eq.~\eqref{eq_ReY_hot} in the main text.

\section{Derivation of coplanar capacitances}
\label{app:coplanar}

We here derive the capacitances of finite size coplanar capacitors, as used in the main text. This calculation is a generalization of the result obtained in Ref.~\cite{Riwar2018}, where the coplanar capacitors were semi-infinite. We start from infinitesimally thin plates in $z$-direction (positioned at $z=0$), infinitely large in $x$ and with finite size in $y$-direction. The distance of the two plates is given by $D$ and their length as $L$ (see also Fig.~\ref{fig_SSp_model} from the main text). We choose to set the origin of $y$ in the middle of the two plates, such that there is a left plate for $-L-D/2<y<-D/2$ and a right plate for $D/2<y<D/2+L$. This problem can be mapped to a parallel plate capacitor through the conformal map,
\begin{equation}
\zeta=K\int^{\tau}d\omega\frac{1}{\sqrt{\omega^{2}-1}\sqrt{\left(l+1\right)^{2}-\omega^{2}}}
\end{equation}
where $\tau=2(x+iz)/D$ is the target space of the coplanar capacitors, and $\zeta$ is the map to the space of the parallel plate capacitor, with the prefactor $K$ to be determined below. We defined furthermore $l=2L/D$. We identify the integral as the incomplete elliptic integral of the first kind, defined as
\begin{equation}
F\left(\frac{\tau}{l+1};\left(l+1\right)\right)=\int_{0}^{\frac{\tau}{l+1}}dt\frac{1}{\sqrt{1-t^{2}}\sqrt{1-\left(l+1\right)^{2}t^{2}}}\ ,
\end{equation}
such that
\begin{equation}
\zeta\left(\tau\right)=iKF\left(\frac{\tau}{l+1};\left(l+1\right)\right)\ .
\end{equation}
The elliptic integral has the following asymptotic behaviours. For $l\gg 1$,
\begin{equation}
F\left(\frac{\tau}{l+1};\left(l+1\right)\right)\rightarrow i\frac{\pi}{2}\frac{1}{\left(l+1\right)}\ ,
\end{equation}
whereas for $l\ll 1$,
\begin{equation}
F\left(\frac{\tau}{l+1};\left(l+1\right)\right)\approx\frac{1}{2}\ln\left(\frac{1}{L}\right)\ .
\end{equation}
We now need to fix the prefactor $K$. We do so by demanding that the point $y+iz=D/2$, that is, $\tau=1$, on the coplanar capacitor corresponds to the point $\zeta=i/2$ in the parallel capacitor space
\begin{equation}
\zeta(\tau=1)\overset{!}{=}\frac{i}{2}\ ,
\end{equation}
resulting in
\begin{equation}
K=\frac{1}{2F\left(\frac{1}{l+1};\left(l+1\right)\right)}\ .
\end{equation}
Eventually, in order to compute the capacitance, we need to know the surface charge on the capacitor plates. We find them as follows. We first solve the parallel plate capacitor in $\zeta$-space for a given voltage difference. In this space, the solution of the potential is simply
\begin{equation}
\varphi'(\zeta)=-V\text{Im}\left[\zeta\right]\ .
\end{equation}
assuming that the parallel plates are situated at $\text{Im}\,\zeta=-1/2$ and $\text{Im}\,\zeta=1/2$.
This is translated to the coplanar capacitor in $\tau$-space as $\varphi(\tau)=\varphi'(\zeta(\tau))$. The surface charge $\sigma$ is then given in the standard way as the normal component of the electric field,
\begin{equation}
\sigma(y)=\epsilon_0 E_n(y)=-\epsilon_0\left.\lim_{\delta\rightarrow 0}\frac{\varphi(\tau+i\delta)-\varphi(\tau)}{\delta}\right|_{\tau=2y/D}\ .
\end{equation}
with $\epsilon_0$ the vacuum permittivity. We receive
\begin{align}
\sigma(y)&=\frac{\epsilon_0 V}{2F\left(\frac{1}{\frac{2L}{D}+1};\frac{2L}{D}+1\right)} \\ & \times \frac{1}{\sqrt{y^2-\left(\frac{D}{2}\right)^2}\sqrt{\left(\frac{D}{2}+L\right)^2-y^2}}\,. \nonumber
\end{align}
In the limit of infinitely large capacitor plates, $L\gg D/2$, we recover our result from Ref.~\cite{Riwar2018}
\begin{equation}
\sigma(y)\approx\frac{\epsilon_0 V}{\pi}\frac{1}{\sqrt{y^2-\left(\frac{D}{2}\right)^2}}\ .
\end{equation}
This is the result we need to estimate the transmon capacitance $C$ and the cross capacitance $C_0$ for large traps. 
The opposite limit of $L\ll D/2$ can be used to estimate the cross capacitance $C_0$ of small traps, where
\begin{equation}
\sigma(y)\approx\frac{\epsilon_0 V}{2\ln\left(\frac{D}{2L}\right)}\frac{1}{\sqrt{y-\frac{D}{2}}\sqrt{\frac{D}{2}+L-y}}\ .
\end{equation}

A coplanar capacitance $C_\text{co}$ can be found in the standard way, by taking the ratio between the total charge stored in the capacitor and the voltage difference. In order to regularize the total charge, we introduce the finite (but large) plate width $W$ in $x$-direction, such that
\begin{equation}
C_\text{co}=\frac{Q_\text{co}}{V}=\frac{\int_0^Wdt\int_{\frac{D}{2}}^{L+\frac{D}{2}}dy\,\sigma(y)}{V}
\end{equation}
For large plates we thus receive
\begin{equation}\label{eq:CcoL}
C_\text{co}\approx \frac{\epsilon_0}{\pi}W\ln\left(\frac{4L}{D}\right)\ .
\end{equation}
For small plates, the capacitance is
\begin{equation}\label{eq:CcoS}
C_\text{co}\approx\frac{\pi}{2}\epsilon_0\frac{W}{\ln\left(\frac{D}{2L}\right)}\ .
\end{equation}
Note that this result does not precisely correspond to the cross conductance $C_0$ for small traps. Namely, in this appendix we consider symmetric coplanar capacitors. The actual setup treated in the main text is in contrast an asymmetric coplanar capacitor, where charges are separated between a large transmon plate and a small trap plate. However, we note that the capacitance of both the small and large (symmetric) coplanar capacitors provide the same scaling, differing only up to logarithmic factors. We can likewise see when doing the spatial integral exactly, that even when crossing over from one regime to the other there appear no additional scaling factors other than $W$. We have therefore no reason to expect that the scaling will be different for the asymmetric case, both on physical grounds as well as from the above scaling argument.

Finally, let us comment on the assumption $C\sim C_0$ used in Sec.~\ref{sec_dissipation}. Equation (\ref{eq:CcoL}) applies to both $C$ and $C_0$ if, for the latter, $D+a \ll L-a$ and the substitutions $L\to L-a$ and $D\to D+a$ are made; then we immediately see that $C/C_0 \gtrsim 1$. In the opposite regime $D+a \gg L-a$, we must use Eq.~(\ref{eq:CcoS}) for $C_0$ (again with the above substitutions); realistically, even for long plates at a typical distance ($L=500\,\mu$m, $D=30\,\mu$m) and a very small trap $L-a = 1\,\mu$m), we estimate $C/C_0 \simeq 4.7$, threfore justifying the assumption $C\sim C_0$.

\bibliographystyle{apsrev4-1}
\bibliography{biblio_qp_trap}

\end{document}